\documentclass[10pt,twocolumn,letterpaper]{article}
\usepackage{wacv}              

\usepackage{graphicx}
\usepackage{amsmath}
\usepackage{amssymb}
\usepackage{booktabs}
\usepackage{colortbl}    
\usepackage{soul}        
\usepackage{xcolor}      
\definecolor{hlcolor}{rgb}{1.0, 1.0, 0.5}  

\usepackage[pagebackref,breaklinks,colorlinks]{hyperref}

\usepackage[capitalize]{cleveref}
\crefname{section}{Sec.}{Secs.}
\Crefname{section}{Section}{Sections}
\Crefname{table}{Table}{Tables}
\crefname{table}{Tab.}{Tabs.}

\def\wacvPaperID{1063} 
\def\confName{WACV}
\def\confYear{2025}

\begin{document}

\title{DiffuPT: Class Imbalance Mitigation for Glaucoma Detection via Diffusion Based Generation and Model Pretraining}

\author{
    Youssof Nawar\textsuperscript{1} \quad
    Nouran Soliman\textsuperscript{1} \quad
    Moustafa Wassel\textsuperscript{1} \quad
    Mohamed ElHabebe\textsuperscript{1} \\
    Noha Adly\textsuperscript{1,2} \quad
    Marwan Torki\textsuperscript{1,2} \quad
    Ahmed Elmassry\textsuperscript{2} \quad
    Islam Ahmed\textsuperscript{2} \\
    \\
    \textsuperscript{1}Applied Innovation Center, MCIT, Egypt \\
    \textsuperscript{2}Alexandria University, Egypt \\
    {\tt\small y.nawar@aic.gov.eg, n.ehab@aic.gov.eg, mostafa.wassel.dio@gmail.com, elhabebem5@gmail.com,} \\
    {\tt\small nadly@aic.gov.eg, mtorki@alexu.edu.eg, ahmad.elmassry@gmail.com, islamshereen@gmail.com}
}

\maketitle

\begin{abstract}
Glaucoma is a progressive optic neuropathy characterized by structural damage to the optic nerve head and functional changes in the visual field. Detecting glaucoma early is crucial to preventing loss of eyesight. However, medical datasets often suffer from class imbalances, making detection more difficult for deep-learning algorithms. We use a generative-based framework to enhance glaucoma diagnosis, specifically addressing class imbalance through synthetic data generation. In addition, we collected the largest national dataset for glaucoma detection to support our study. The imbalance between normal and glaucomatous cases leads to performance degradation of classifier models. We created a more robust classifier training process by combining our proposed framework leveraging \textbf{diffusion models} with a pretraining approach.
This training process results in a better-performing classifier. The proposed approach shows promising results in improving the harmonic mean ``sensitivity and specificity'' and AUC for the roc for the glaucoma classifier. We report an improvement in the harmonic mean metric from 89.09\% to 92.59\%  on the test set of our Egyptian dataset. We examine our method against other methods to overcome imbalance through extensive experiments. We report similar improvements on the AIROGS dataset. This study highlights that diffusion-based generation can be important in tackling class imbalances in medical datasets to improve diagnostic performance.
\end{abstract}

\section{Introduction}
\label{sec:intro}

\textbf{Glaucoma condition} is an eye disease primarily affecting peripheral vision. If left untreated, it can lead to blindness \cite{hu2014patients}. Therefore, regular examinations targeting specific age groups and health conditions are crucial for early detection and intervention to prevent further eye damage \cite{tatham2014strategies}. Recent advancements in computer vision, particularly deep learning algorithms, have been employed to facilitate the detection of glaucomatous eyes.

In this study, we explore various approaches to address glaucoma's low prevalence, estimated to be approximately 3.54\% according to Tham et al. \cite{tham2014global}. This prevalence is also evident in all the available datasets.

\textbf{Class imbalance} is a persistent issue in datasets, especially in medical datasets, where data collection and annotation can be costly and diseases may be underrepresented. Data sampling techniques can be employed  \cite{van2007experimental} where classes are oversampled or undersampled accordingly. Cost-sensitive learning employs a penalty matrix to penalize minority classes loss more  \cite{krawczyk2016learning, sun2007cost, thai2010cost}. Another technique is decoupled representation learning, where feature extraction is decoupled from classifier training \cite{kang2019decoupling}. 
\newline
Standard image augmentation employs simple transformations like rotations, flips, and brightness adjustments, providing computational efficiency and ease of implementation. This simplicity facilitates widespread applicability across diverse datasets and models, improving robustness and generalization. However, these basic techniques may fail to capture the intricate patterns present in some datasets. Diagnostic tasks demand a deeper understanding of higher-level semantics. These tasks require more sophisticated augmentation strategies involving domain-specific knowledge to optimize model performance.

In recent works, generative vision models  \cite{goodfellow2014generative,ho2020denoising,rombach2022high} have emerged generating high-quality images however, the use of these models in other computer vision tasks was quite unclear. Several attempts to generate synthetic images to be used in dataset augmentation \cite{skandarani2023gans, perez2017effectiveness}. These attempts were either on simple datasets or didn't improve the performance against other techniques. 

This work aims to introduce a clearer study of generative models in a more intricate domain such as eye fundus. We aim to utilize these generated images for the glaucoma classification task. We propose a new training scheme that can outperform all mentioned techniques in addressing the class imbalance problem. 

In addition, we introduce our Egyptian dataset for Glaucoma Detection (\textbf{GlaucomaEgy}). While existing datasets are available for glaucoma detection like AIROGS \cite{de2023airogs}, LAG \cite{li2019attention}, and REFUGE \cite{orlando2020refuge}. There exists ethnic variation in fundus images which refers to differences in the appearance of the retina among individuals of different ethnic backgrounds \cite{jacoba2023ethnic, li2013racial}. These variations significantly affect glaucoma detection \cite{li2013racial, tielsch1991racial}. To overcome these variations, we collected more than 37K samples, with roughly 10 percent having glaucomatous eyes. That was through a countrywide effort that targeted local hospitals from all over the nation. In section \ref{sec:Dataset}, we will go over the dataset in details.  

\textbf{Our proposed framework} aims to leverage the generative model's ability to capture the complex distribution of the data. In addition to its ability to clearly distinguish the class-specific attributes to generate data that can be used in a pre-training methodology to improve the classifier results. This can be broken down into several steps. Initially, we train a generative model to generate high-quality images for both glaucomatous and non-glaucomatous fundus images without any additional annotation. Next, we use the newly generated data as a pretraining for our desired classifier based on baseline experiments. Finally, we Fine-tune the classifier on the real dataset to improve the classification results. More details on experimental results and the final framework will be discussed in sections \ref{sec:Proposed Approach} and \ref{sec:Experiments}.

To summarize, our main contributions are:
\begin{itemize}
\item We introduce the largest national dataset for glaucoma detection (\textbf{GlaucomaEgy}), which has helped us bridge the gap between available open-source data and our national data.
\vspace{-5pt}
\item Introducing a framework that utilizes a generative model for enhancing other computer vision tasks i.e. glaucoma classification. Our framework requires no additional annotations for the generated data.  
\vspace{-5pt}
\item Comparison between diffusion-based and GAN-based generation specifically between conditional diffusion models and image-to-image translation GANs.
\vspace{-5pt}
\item Comparison between our proposed framework for training and current techniques used for mitigating class imbalances through extensive experiments.
\end{itemize}

\section{Related Work}
\begin{figure}[t]
    \centering
    \includegraphics[width = 1.0 \linewidth]{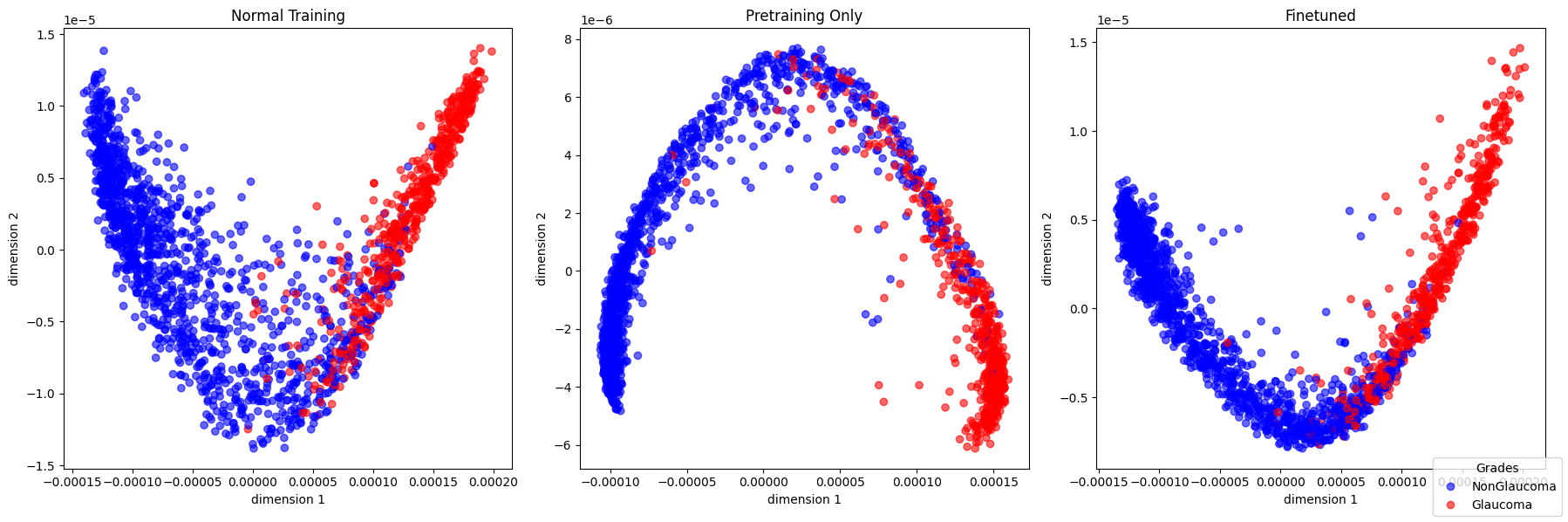} 
    \caption{Analysis of the embeddings on different training techniques. Normal Training shows high variance and low bias (less overlap between glaucoma and non-glaucoma classes. Pre-training alone shows much more stability with lower variance but with higher overlap. The fine-tuned model shows the best of both worlds, a robust classifier with lower variance and lower bias.}
    \label{fig:Latent Space}
\end{figure}
\subsection{Deep Learning in Glaucoma}
Deep learning methods have demonstrated remarkable results across various tasks in fundus imaging analysis\cite{elhabebe2024dr10k}. Multiple CNN architectures, such as those discussed in \cite{li2018efficacy,deperlioglu2022explainable,kucur2018deep}, have been employed for the early detection of Glaucomatous eyes and the segmentation of fundus images \cite{zhao2019weakly}. These architectures have shown excellent performance in extracting relevant features for classification.

In recent developments, Transformers \cite{vaswani2017attention} have been introduced into vision tasks, achieving impressive results. The Vision Transformer (ViT) \cite{dosovitskiy2020image} is a deep learning architecture that utilizes self-attention mechanisms to capture global characteristics in images. Various Vision Transformer architectures, as highlighted in \cite{wassel2022vision}, have been applied to glaucoma detection surpassing the performance of CNNs.


\subsection{Generative Models}
\subsubsection{Generative Adversarial Networks (GANs)}

GANs \cite{goodfellow2014generative} represented a breakthrough in the field of synthetic data generation and especially image generation. The evolution of GAN-based generators depends on adversarial training, where a discriminator is trained to differentiate between real and fake data minimizing a discrimination loss while the generator tries to maximize it. Recently, GANs were used in image-to-image translation problems to transform domain-specific features from one image to another. However, these changes should occur without changing the perceptual features of the original image. Later, CycleGAN \cite{zhu2017unpaired} was introduced to perform unpaired image-to-image translation. The main idea in CycleGAN is to use two generators, one to convert from the first domain to the second, while the other converts from the second domain back to the first. The two generators are optimized collaboratively using the cycle consistency loss in addition to the traditional adversarial loss with the discriminator of each one. The cycle consistency loss is represented by  equation~\ref{Ganeqn}:
\begin{equation}
\label{Ganeqn}
\mathcal{L}_{CYC} := \left\| I_1 - G_{1}(G_{2}(I_1)) \right\|_1
\end{equation}
\begin{figure*}[th]
    \centering
    \includegraphics[width=0.9\linewidth]{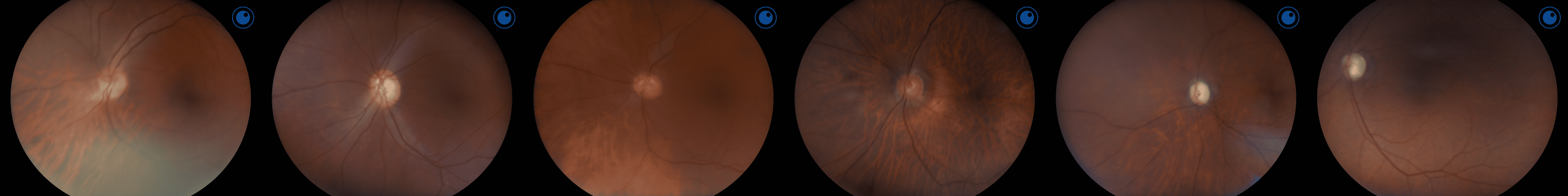} 
    \caption{Samples of the collected dataset. The two images on the left are from grade 0 (non-glaucomatous) samples. Both middle images are from grade 1 (glaucoma suspect). The two right images are from grade 2 (glaucoma).}
    \label{fig:Dataset samples}
\end{figure*}
where $I_1$ is an image from the first domain, $G_{1}$ is the generator that converts from the second domain to the first domain, and $G_{2}$ is the generator that converts from the first domain to the second domain.
In Section \ref{sec:Experiments},  we go through the use of multiple GANs architectures i.e. CycleGAN, StarGANV2\cite{choi2020stargan}, MW-GAN\cite{cao2019multi}, and the most recent UNSB \cite{kim2023unpaired} in the generation of fundus images.

\subsubsection{Diffusion Generative Approaches}
Denoising diffusion probabilistic models (DDPM) learn to progressively remove noise from
random initial vectors into high-resolution images through an iterative sampling process \cite{ho2020denoising,song2020score}. The learned network $\epsilon_\theta(x, t)$ in the diffusion model tries to estimate the noise $\epsilon_t$ that was added at a previous step $t - 1$. We use signal $y$ for conditional generation \cite{ho2022classifier}. The loss function to be optimized is as follows:
\begin{equation}
\label{eq: diff_loss}
L(\theta) = \mathbb{E}_{(t \sim U(1,T), \epsilon_t \sim \mathcal{N}(0, I)}\left[ w(t) \cdot \|\epsilon_t - \epsilon_\theta(z_t; t, y)\|^2 \right]
\end{equation}
Where $y$ represents a conditioning signal, such as text or a class or an image, $w(t)$ is a weighting function and $z_t$ represents the noisy image at timestep $t$. The architecture used for $\epsilon_\theta$ is typically a U-net architecture. Cross-attention can be used for text or class conditioning.

During sampling, we can generate samples by sampling noisy image $x_T$ $\sim \mathcal{N} (0, I)$ and $y$ the desired class. We update the noisy image using the following:
\begin{equation}
\label{eq:sample}
x_{t-1} = \ \frac{1}{\sqrt{\alpha_t}} \left( x_t - \frac{\sqrt{1-\alpha_t}} {1-\bar{\alpha}_t} \right) \epsilon_\theta(x_t, t , y) + \sigma_t.z
\end{equation}
where $\alpha_t$ = $1 - \beta_t$ and  $\bar{\alpha}_t = \prod_{s=1}^{T} \alpha_s$ and $\beta_t$ is linearly scheduled to control the mean on the noise added to the original image. $z$ $\sim \mathcal{N} (0, I)$ and $\sigma_t$ are used to control the stochasticity of sampling.

\subsection{Class imbalance}
Data imbalance is a common problem facing most datasets especially in the medical field. Data balancing techniques has been employed to mitigate this problem, researchers have proposed to re-sample the imbalanced dataset to achieve a more balanced data distribution. These methods include over-sampling for the minority classes (deliberately choosing the same sample more than once), under-sampling for the majority classes (by removing data)\cite{van2007experimental}, and class-balanced sampling \cite{mahajan2018exploring} based on the number of samples for each class. The Synthetic Minority Oversampling Technique (SMOTE) \cite{chawla2002smote} generates artificial minority samples by interpolating between existing minority samples and their closest neighbors.

Algorithmic techniques can be employed to address class imbalance without altering the distribution of the training data. These methods involve adjusting the learning or decision process to enhance the importance of a specific class\cite{lin2017focal}. This can be achieved by modifying algorithms to incorporate class penalties or weights or shifting the decision threshold to mitigate bias towards the other classes.

Cost-sensitive learning is another approach where a cost matrix allocates penalties to each class. Increasing the cost associated with the minority group enhances its significance and reduces the probability of incorrect classifications for instances from this group \cite{krawczyk2016learning}.

Another technique introduced to overcome class imbalances was decoupling representation and classifier for long-tailed recognition \cite{kang2019decoupling} where learning features are disjointed from learning a classifier to the obtained features. This multi-stage training gives more generalizable representations. This can improve classification by re-balancing the classifiers without the need for carefully designed losses.
\vspace{-8pt}
\section{Datasets}
\label{sec:Dataset}
\subsection{GlaucomaEgy Dataset} 
We collected the data from 20,445 patients. Each patient contributes either left or right fundus images or both. A total of 39,449 fundus images were collected of which 37,253 were gradable. The dataset is collected at diverse sites and synced to a database system for later grading by twelve glaucoma-specialized ophthalmologists. The final dataset consists of 33,633 fundus images of grade 0 (nonglaucoma), 2,081 grade 1 (suspicious glaucoma), and 1,539 grade 2 (glaucoma). We show samples from the dataset in Figure \ref{fig:Dataset samples}. Building on the study by \cite{elmassry2023prevalence} on fundus imaging and the prevalence of another macular disease in Egypt, we present the results of our demographic analysis in Table \ref{table:demographic}.

\begin{table}[htbp]
\centering
\resizebox{\linewidth}{!}{
\begin{tabular}{@{}cccccc@{}}
\toprule
  &  & Grade 0 & Grade 1 & Grade 2 & Total \\
\midrule
 & 0 - 18  &  174  & 19  & 28 & 221  \\
 & 18 - 30   &  3709  & 159  & 85  & 3953  \\
 Age & 30 - 45   &  10408  & 443  & 207  & 11058 \\
 & 45 - 60   &  11286  & 615  & 412  & 12313  \\
 & 60 +   &  8056  & 845 & 807  & 9708  \\
\midrule  
Gender & Male  &  12071  & 1009  & 953  & 14033  \\
& Female  &  21562  & 1072  & 586  & 23220  \\
\midrule  
 Diabetic Status & True  &  5596  & 436  & 309  & 6341  \\
 & False  & 28037  & 1645  & 1230 & 30912  \\
\midrule
 Type & Left fundus  &  16916  & 994  & 772  & 18682  \\
 & Right fundus  & 16717  & 1087  & 767 & 18571  \\
\bottomrule
\end{tabular}}
\caption{Demographic analysis for GlaucomaEgy.}
\label{table:demographic}
\end{table}
\section*{Grading Mismatch analysis}

We rely on an established process for grading our data, which involves the annotations of two experts. In case of any mismatch between the two graders, the samples are sent to a more experienced specialist (adjudicator). This is the same procedure used for grading other datasets such as the AIROGS dataset\cite{de2023airogs}. The grading can be split into two main phases, quality mismatch and grading mismatch. Quality mismatch implies that one of the graders identified the image as gradable while the other didn't. Grading mismatch implies conflict in grading a fundus image. In any case of a mismatch, an experienced adjudicator resolves the tie.
\newline
We discovered some disparity in grading amongst graders, as illustrated in Table \ref{table:Mismatch Analysis}. The mismatches between grades 1 and 2 are less significant, as they will later be combined into a single class (referable glaucoma). Quality mismatches are also relevant as the quality of the eye fundus image is important in whether it is gradable or not. We observed that for grade 0, a total of 3,818 samples were confused among the two graders, and 1,441 were due to quality mismatch. 2,358 had at least one agreement with one of the graders. For the remaining 19 samples, the adjudicator completely disagreed with both graders. Grade 1 and grade 2 analyses are shown in Table \ref{table:Mismatch Analysis}.
\begin{table}[htbp]
\centering
 \resizebox{\linewidth}{!}{
\begin{tabular}{@{}ccccc@{}}
\toprule
 & Disagreement with& Grade 0 & Grade 1 & Grade 2\\
\midrule
Quality mismatch && 1,441 (4.28\%)  &  204 (9.8\%)  & 194 (12.61\%)\\
\midrule
Grading Mismatch & One Grader& 2,358 (7.01\%)  & 849 (40.8\%) &  639 (41.52\%) \\
 &Both Graders & 19 (0.06\%)   & 57 (2.74\%)  &  49 (3.18\%) \\
Total && 3,818 (11.35\%)  & 1,110 (53.34\%)  &  882 (57.31\%) \\
\bottomrule
\end{tabular}}
\caption{Quality and Grade mismatch per grade. The final grade is given by the adjudicator.}
\label{table:Mismatch Analysis}
\end{table}


\section*{Dataset Split}

 We used a training set of 31,047. We refer to grade 0 as nonglaucoma and combine grades 1 and 2 as referable glaucoma. The training set is divided into 28,418 non-glaucoma and 2,629 glaucoma. Our Validation set contains 4,343 fundus images of which 471 are glaucoma and 3,872 are nonglaucoma. Following the works of \cite{de2023airogs,li2019attention,wassel2022vision}, we choose to set the test set distribution at around 28\% with 521 glaucoma samples and 1,342 nonglaucoma images. The split is shown below in Table \ref{table:Dataset}.
 \begin{table}[htbp]
\centering
 \resizebox{\linewidth}{!}{
\begin{tabular}{@{}cccc@{}}
\toprule
 & Train & Validation & Test\\
\midrule
Glaucoma & 2,629 (8.47\%)  & 471 (10.85\%) &  521 (27.97\%)\\
Non-Glaucoma & 28,418 (91.53\%) &  3,872 (89.15\%) & 1,342 (72.03\%)\\
Total  & 31,047 & 4,343 & 1,863\\
\bottomrule
\end{tabular}}
\caption{Dataset Distribution}
\label{table:Dataset}
\end{table}



\begin{figure*}[th]
    \centering
    \includegraphics[width=0.75\linewidth]{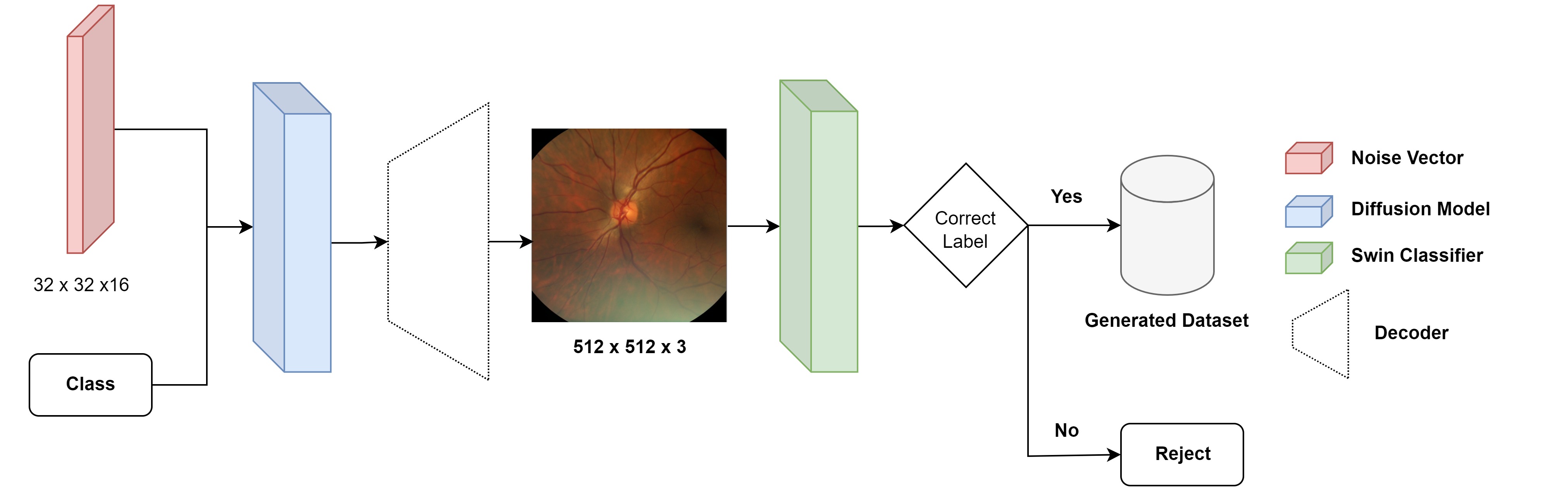} 
    \caption{Framework for generating data using LDM. We show how to sample a generated dataset using LDM. Our method involves a filtering stage based on the classification output of the baseline classifier.}
    \label{fig:Diffusion process generation}
\end{figure*}
\begin{figure}
    \centering
    \includegraphics[ width = 0.8 \linewidth]{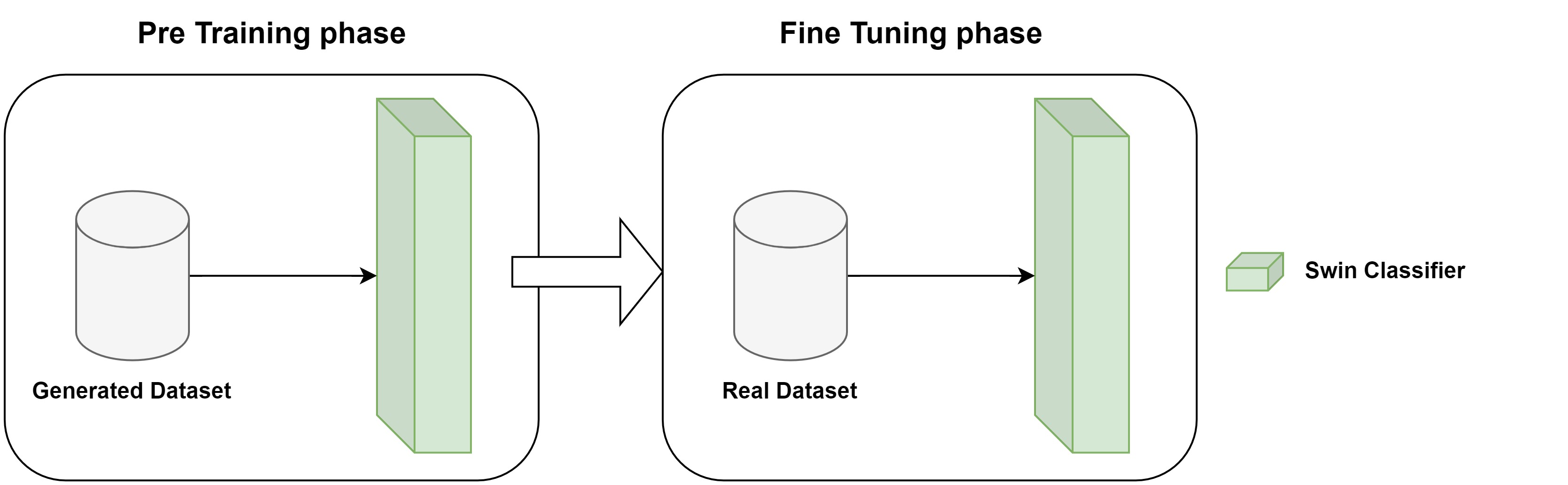}
    \caption{Training procedure used for obtaining the final classifier.}
    \label{fig:finetune}
\end{figure}
\subsection{AIROGS}
\textbf{AIROGS} \cite{de2023airogs} dataset is a collection of 112,732  fundus images collected from 500 different sites. The dataset was graded by thirty carefully chosen optometrists and ophthalmologists. Based on predetermined glaucomatous indications, each eye was graded as referable glaucoma (RG), non-referable glaucoma (NRG), or ungradable (U). Each eye underwent evaluation by two independent graders. If a case of a mismatch arises, the image is assessed by a glaucoma specialist (adjudicator). The dataset was divided into training and test sets, the former consists of 101,442 images while the latter
made up of 11,290 images, ensuring that data from patients in the training set was not in the test set. The training set consists of 98,172 NRG and 3,270 RG. 
\newline
The test dataset is held out, so we evaluate a subset of the training set. We used 99,267 samples for training. Both validation and test were held at 1,106 and 1,069, respectively, to hold the same distribution as the test set in the AIROGS challenge at 14\% referred glaucoma (RG).
\section{Proposed Approach}
\label{sec:Proposed Approach}

We aim to find $f_\theta(x)$ that is parameterized by network parameters $\theta$ that can classify any given image $x$ to its correct label $y$. 
Let $\mathcal{D} = \{ (\mathbf{x}_i, y_i) \}_{i=1}^N$ be a dataset, where $\mathbf{x}_i$ represents the input image and $y_i \in \{+1, -1\}$ denotes the class labels. Let $N_{positive}$ and $N_{negative}$ be the number of referable glaucoma and non-glaucoma class samples, respectively. We have:

\[
N_{negative} \gg N_{positive}
\]
This class imbalance affects $f_\theta$ training. 
We aim to synthesize a new dataset $\hat{\mathcal{D}}$ through a generative model that doesn't suffer from the same class imbalance. This dataset $\hat{\mathcal{D}}$  is to be sampled from the surrogate posterior $p(x|y)$ estimated by the generative model.
The final dataset should have:
\[
\hat{N} > N
\]
\[
\hat{N_{positive}} \approx \hat{N_{negative}}
\] Our final classifier will be finetuned on the imbalanced dataset $\mathcal{D}$ at a lower learning rate.

Pretraining involves training the neural network on a large, generic dataset to learn useful representations. This process can be formalized as follows:
 Given the pretraining dataset $\hat{\mathcal{D}}$ the network parameters $\theta$
 are optimized to minimize the pretraining loss function $\mathcal{L}_{\text{pretrain}}$:
\[
\theta_{\text{pretrained}} = \arg \min_{\theta} \mathcal{L}_{\text{pretrain}}(\hat{\mathcal{D}}, \theta)
\]

Fine-tuning adapts the pre-trained model to a specific target task using our real dataset with distribution $p(x)$, task-specific dataset $\mathcal{D}$. Starting from the pre-trained parameters $\theta_{\text{pretrained}}$, the parameters are further optimized to minimize the fine-tuning loss function $\mathcal{L}_{\text{finetune}}$ on the fine-tuning dataset $\mathcal{D}$:
\[
\theta_{\text{finetuned}} = \arg \min_{\theta} \mathcal{L}_{\text{finetune}}(\mathcal{D}, \theta)
\]

\label{sec:DiffuPT}
\subsection{DiffuPT}
Our method initially involves the use of diffusion models. We chose the Latent Diffusion Model \cite{rombach2022high} (LDM) for sample generation. In LDM, the diffusion process operates in latent space, improving efficiency, speed, and sample quality.
LDM is composed of two main modules: a denoising diffusion probabilistic model and an auto-encoder.
The loss we used for training the diffusion model denoted as $\mathcal{L}_{LDM}$, is defined as in the following.
\begin{equation}
\label{eq:LDM loss}
\mathcal{L}_{LDM} := \mathbb{E}_{(\varepsilon(x), y, \epsilon \sim \mathcal{N}(0,1), t)} \left\| \epsilon - \epsilon_{\theta}(z_t , t, \tau_{\theta}(y)) \right\|_2^2
\end{equation}
Where $x$ represents the original image, $y$ is the class label (0 for non-glaucoma, 1 for glaucoma), $\epsilon$ is Gaussian noise,$z_t$ is the noised latent vector, and $\theta$ denotes model parameters. The estimated noise is the output of the model $\epsilon_{\theta}(z_t, t, \tau_{\theta}(y))$, where $\tau_{\theta}(y)$ is a class embedder with learnable parameters.

The auto-encoder in our diffusion model comprises an encoder and a decoder. The encoder compresses an image $I$ to a lower-dimensional latent space vector $z$:
\[
    z = \text{Encoder}(I)
\]
where $z$ is a 32 x 32 x 16 vector, the Decoder reconstructs the original image from the latent space retaining the 512 x 512 x 3 image:
\[
\hat{I} = \text{Decoder}(z)
\]
\newline
The encoder-decoder was trained on a collection of AIROGS and GlaucomaEgy datasets on LPIPS  (Learned Perceptual Image Patch Similarity) \cite{zhang2018unreasonable} loss. The objective is to initially identify a perceptually equivalent space that is computationally more suitable. We need to ensure the fundus image can be successfully reconstructed (SSIM = 92.53).

For the sampling process, We used DDIM \cite{song2020denoising} sampler at 250 steps for faster generation.

\subsection{Procedure}
Our final framework, DiffuPT can be broken down into three stages, Initially, we train a diffusion model for generating specific domain images. Then, a generation stage will be generated in which we sample $z$ $\sim \mathcal{N} (0, I)$ and target class $y$. The diffusion model denoises the input image to generate a suitable vector. The latter is decoded to generate our desired image which is then filtered using our already trained baseline classifier to determine whether it is from the desired class or not as shown in Figure \ref{fig:Diffusion process generation}.
\newline
Finally, the final generated dataset with the desired distribution (balanced classes) is later used for the classifier model pretraining. Subsequently the finetuning is performed on the real dataset. The training scheme to obtain the final classifier can be shown in Figure \ref{fig:finetune}. 
\newline
This training mechanism helped reduce the variance of the data embedding in the latent space. Recent work by \cite{shwartz2024simplifying} showed that the variance varies depending on the distribution of the dataset which is consistent with our results shown in Figure \ref{fig:Latent Space}. The pretraining with a balanced dataset leads to more stable (low variance) classifier embeddings, however, has a higher bias (overlap) between both classes due to the noise in the generated samples. Thus fine-tuning using the real dataset leads to the best classifier.
\section{Experiments and results}
\label{sec:Experiments}
In this section, we present detailed results of different training approaches. Initially, we present the baseline results to determine the best architecture. Secondly, we contrast seven techniques for fundus image generation. Thirdly, we show the results of generation-based augmentation. Finally, we show the effect of DiffuPT and assess generalization on the AIROGS dataset.
\newline
We explored various approaches to address the imbalance, including weighted cross-entropy loss, weighted sampling based on class weights, multi-stage training by decoupling representation, classifier learning, and combinations of these strategies. We show that the DiffuPT technique outperforms all the above-mentioned techniques.

\subsection{Baseline Classifiers}
We trained two convolution-based and three vision transformer-based classifiers to choose the best-performing architecture. The results in Table \ref{table:standalone_Test} show the test results on our national dataset.
We trained our national dataset at a learning rate (lr) 1e-4 and Adam optimizer for 60k iterations and batch size 64 using four Tesla V100s.





\begin{table}[htbp]
\centering
 \resizebox{\linewidth}{!}{
\begin{tabular}{@{}ccccc@{}}
\toprule
Model & Sensitivity & Specificity & AUC & Harmonic Mean\\
\midrule		
Swin large\cite{liu2021swin} & 83.69 & 95.23 & \textbf{97.62} & \textbf{89.09}\\
Cait XXS-36\cite{touvron2021going} & 75.61 & \textbf{96.8} & 96.8 & 84.9\\
ViT Large \cite{dosovitskiy2020image}  & 82.9 & 90.5 & 93.2 & 86.53\\
EfficientNet B5\cite{tan2019efficientnet}  & 82.92 & 91.57 & 94.91 & 87.03\\
ResNet 101\cite{he2016deep}  & \textbf{87.8} & 87.36  & 87.58 & 87.58\\
\bottomrule
\end{tabular}}
\caption{Results of different architectures.}
\label{table:standalone_Test}
\end{table}

We observed that the sensitivity is low because of the significant class imbalance. We determined that increasing the harmonic mean\footnote{Harmonic mean computed between sensitivity and specificity.} should be our goal. SWIN transformer \cite{liu2021swin} had the best harmonic mean at 89.09\% while retaining the highest AUC than the other architectures. Therefore all upcoming experiments are performed with SWIN as its backbone and will be used as our baseline classifier for future experiments.
\begin{figure*}[ht]
  \centering
    \includegraphics[width=0.9\linewidth]{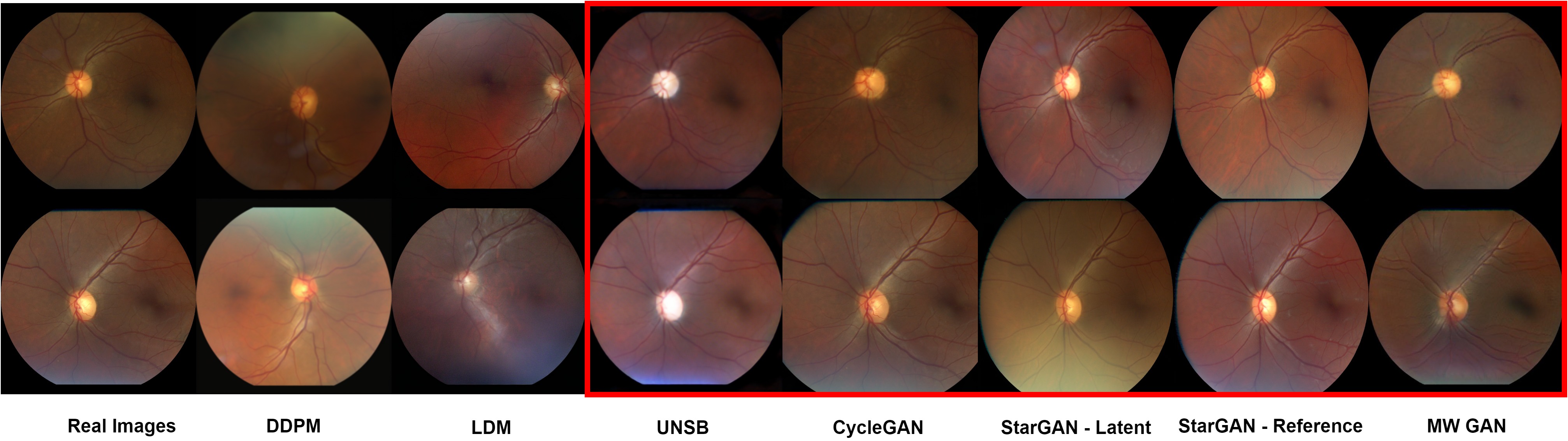} 
      \caption{Samples from different generation methods. The left images show nonglaucoma samples from the dataset. Then we show the results of conditional generation from the DDPM and LDM. The bounded images are from GAN-based image-to-image translation from real nonglaucoma images to fake glaucoma using various methods. }
    \label{fig:Samples}
  \hfill
\end{figure*}

\subsection{Generation for Augmentation}
In this section, we represent the results of the generative approaches to augment the less-represented glaucoma class. We investigated many strategies for employing generative algorithms to determine the optimal method.
We will consider both diffusion-based and GAN-based approaches.

\subsubsection{GANs}

We implemented CycleGAN as introduced in~\cite{zhu2017unpaired}, the discriminator networks are PatchGANs \cite{ledig2017photo} and the generator network was a U-Net~\cite{ronneberger2015u} based generator. The input image is resized to 512 x 512 at a batch size of twelve. The model was trained for five days on four V100 GPUs.

The StarGANV2\cite{choi2020stargan} technique includes four networks, which are generator, discriminator, style encoder, and mapping network. We used the trained generator to convert each image from the training set to the other class ten times using either the latent or reference method. We used the same architectures described in \cite{choi2020stargan} for all the networks. We trained the four networks for 100K iterations using a batch size of eight and an image size of 768 x 768. The training lasts for seven days on four V100 GPUs.

The MWGAN\cite{cao2019multi} technique improves upon cycle gan with the help of multi-marginal optimal transport theory, They developed a new dual formulation for better adversarial learning on the unsupervised multi-domain image translation task. We trained the four networks for 100K iterations using a batch size of eight and an image size of 512 x 512. The training lasts for two days on four V100 GPUs.

The UNSB\cite{kim2023unpaired} is a recent technique that combines the GAN training techniques with the Shrodinger bridge training techniques for unpaired image-to-image tasks. We trained the network for 100K iterations using a batch size of 20 samples and an image size of 512 x 512. The training lasts for one day on four V100 GPUs.

\subsubsection{Diffusion Models}
We trained Conditional DDPM \cite{ho2022classifier} for 256 x 256 image generation at a batch size of 16 on four V100s for two days. The diffusion process requires more computations so an image size larger than 256 x 256 would be infeasible.

We trained the latent diffusion model, LDM improves upon DDPM in that the diffusion process is performed in much lower dimensional latent space. This can lead to high image resolution generation and faster sampling. The network consists of an encoder, a diffusion model, and a decoder. The autoencoder was trained and frozen during the training of the LDM. We trained our network on 512 x 512 images for four days on four V100s with a batch size 16 for 120k iterations.

\subsubsection{Generative Techniques Comparisons}
We show samples of the generated data in Figure \ref{fig:Samples}. The GANs are used in image-to-image translation so we show the real non-glaucoma image and the corresponding glaucoma sample for each model. In the diffusion models, we sample conditionally for the glaucoma class. We conducted the Fréchet Inception Distance (FID) \cite{heusel2017gans}, Kernel Inception Distance (KID) \cite{binkowski2018demystifying}, and inception score \cite{salimans2016improved} for quantitative measures. A lower FID and KID score indicates that the generated images resemble the real distribution more. Table \ref{table:generation_quantitative} shows that LDM scored the lowest FID and KID, while the DDPM scored the best inception score. This indicates that the diffusion models are superior to their GAN counterparts. One draw back for the diffusion models is sampling time as it was much higher than the GANs,
\begin{table}[htbp]
\centering
\resizebox{0.9\linewidth}{!}{
\begin{tabular}{@{}c c c c c c@{}}
\toprule
Model & NFE & FID $\downarrow$ & KID $\downarrow$ & IS $\uparrow$ &  Sampling time (s) \\ 
\midrule
Cycle GAN & 1 & 48.63 & 0.029  & 167.6 &  0.011 \\
StarGANV2-L & 1 &  39.76  & 0.015  & 152.57 & 23 \\
StarGANV2-R & 1 &  41.88  & 0.014  & 150.09 & 25  \\ 
MW-GAN & 1 &  119.72  & 0.103 & 157.49 &  0.084 \\
UNSB & 5 &  60.61  & 0.031 & 169.75 &  1.17\\
DDPM & 250 &  37.27  & 0.041 & \textbf{188.39} &  136 \\
LDM & 250 & \textbf{7.94} & \textbf{0.006} & 153.74 &  80 \\
\bottomrule
\end{tabular}}
\caption{Generation results. NFE refers to the number of function evaluations.}
\label{table:generation_quantitative}

\end{table}

\subsubsection{Generation Based Augmentation}

\begin{figure}[htbp]
    \centering
    \includegraphics[width = 1.0 \linewidth]{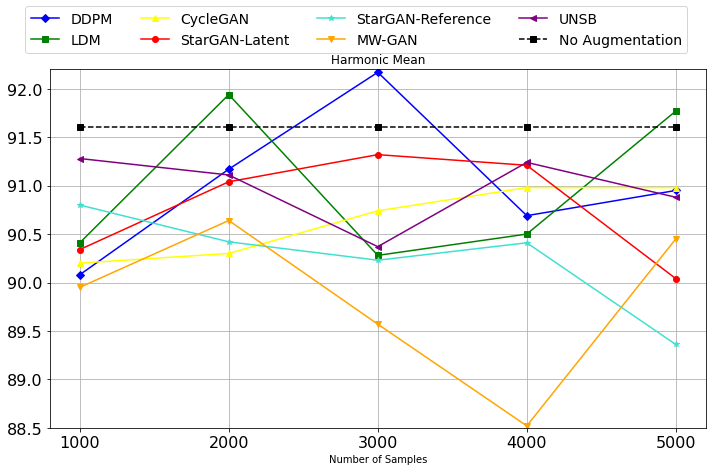} 
    \caption{Results of augmentation using generated samples (Training is performed using weighted samplers).}
    \label{fig:Ablation}
\end{figure}

To study the effect of the augmentation, we sampled glaucoma images and aimed to increase the number of samples iteratively until the results started worsening.
We started with 1,000 up to 5,000 samples. An increase in synthetic samples used deteriorated the classifier's capabilities beyond the 5,000 samples. Figure \ref{fig:Ablation} presents the harmonic mean of each method against the number of augmented samples.

We observed that the augmentation outcomes varied proportionally with the generation quality shown in Table \ref{table:generation_quantitative}. Diffusion Models performed better than the GANs yet again. DDPM had the best harmonic mean surpassing no augmentation at 92.17 followed by the LDM at 91.77. On average, the best augmentation results peaked around the 2000-3000 samples, the same number of glaucoma samples in the training dataset (2629). 

\subsection{DiffuPT}
We opted with the LDM as it had the best generation scores, faster sampling concerning the DDPM, and surpassed normal training in augmentation results.
\newline
DiffuPT can be decomposed into two stages a pretraining stage where we used the generated data for training the classifier. We used the same validation set to determine the best initial weights. The training was performed using four V100 32GB for 100k iterations at lr equal to 1e-4.
Then a fine-tuning phase using the obtained weights. We used the original ill-distributed dataset. The training was performed using four V100 32 GB for 60k iterations at lr equal to 1e-5. We used BCE loss for $\mathcal{L}_{\text{pretrain}}$ and $\mathcal{L}_{\text{finetune}}$.
\newline
As an ablation, we tried different glaucoma to non-glaucoma distributions for pre-training, a total of 96,307 samples were generated of which 43,550 were glaucoma. We tried 5 distributions for pre-training from 30-70 (30\% glaucoma) up to 70-30 (70\% glaucoma). The model was validated against our validation set before fine-tuning. Table \ref{table:Ablation} showed that the 50-50 split performed the better AUC and was slightly behind the 60-40 split in the harmonic mean. The more balanced datasets performed better than the more biased distributions.

\begin{table}[htbp]
\centering
 \resizebox{\linewidth}{!}{
\begin{tabular}{c c c c c}
\toprule
Data Distribution  & Sensitivity & Specificity & AUC & harmonic mean\\
\midrule
30-70    & 91.51 & 90.57  & 96.9 & 91.03\\   
40-60    & 91.72 & 90.29  & 97.13 & 91\\ 
50-50    & \textbf{92.78} & 90.13  & \textbf{97.27} & 91.44\\ 
60-40    & 91.08 & \textbf{91.97} & 97.1 & \textbf{91.52}\\ 
70-30    & 90.45 & 91.94 & 97.11 & 91.19\\ 
\bottomrule
\end{tabular}}
\caption{Ablation on different pretraining dataset distributions for training (without fine-tuning). }
\label{table:Ablation}
\end{table}

\begin{table}[htbp]
\centering
 \resizebox{\linewidth}{!}{
\begin{tabular}{c c c c c}
\toprule
Method  & Sensitivity & Specificity & AUC & harmonic mean\\
\midrule	
Normal Training  & 85.56  & \textbf{94.6} & 97.34  & 89.85  \\ 
Weighted CE   & 89.81  & 93.23 & 90.27 & 91.48 \\ 
Weighted Sampler & 90.45 & 91.58 & 97.22 & 91\\
Weighted CE + Sampler & \textbf{94.26} & 89 & 92.37 & 91.55 \\
Multi-Stage training + Sampler  & 92.57 & 90.34 & 97.13 & 91.44 \\
\midrule
Augmentation method   & 90.23 & 91.61 & 97.12 & 90.91\\
DiffuPT   & 91.72 & 92.43 & \textbf{97.56} & \textbf{92.07} \\
\bottomrule
\end{tabular}}
\caption{Comparison of classification results using different training methods - GlaucomaEgy Validation.}
\label{table:Final_Validation}
\end{table}
We present our final results on the GlacomaEgy data in Tables  \ref{table:Final_Validation} and \ref{table:Final_Test} for validation and test respectively. We compare DiffuPT to normal training without a weighted sampler against training with weighted cross entropy (CE) loss, weighted sampler, generation-based augmentation, and Multi-Stage training. The results indicate that DiffuPT outperforms all its counterparts. The confusion matrices for the augmentation and diffuPT can be found in the appendix (see section results analysis in the appendix). 
\begin{table}[htbp]
\centering
 \resizebox{\linewidth}{!}{
\begin{tabular}{c c c c c}
\toprule
Method  & Sensitivity & Specificity & AUC & harmonic mean\\
\midrule
Normal Training  &  83.69 & \textbf{95.23} & 97.62 & 89.09 \\ 
Weighted CE   & 93.86  & 89.12 & 89.17 & 91.42 \\ 
Weighted Sampler & 95 & 88.45 & 97.47 & 91.61\\
Weighted CE + Sampler & \textbf{97.5} & 83.90 & 90.09 & 90.23 \\
Multi-Stage training + Sampler  & 94.05 & 88.23 & 97.28 & 91.05 \\
\midrule
Augmentation method   & 93.86 & 90.54 & 97.4 & 92.17\\
DiffuPT   & 93.09 & 92.1 & \textbf{98.02} & \textbf{92.59} \\
\bottomrule
\end{tabular}}

\caption{Comparison of classification results using different training methods - GlaucomaEgy Test.}
 \label{table:Final_Test}
\end{table}
\section*{Effect of filtering}
To study the effect of filtering the generated samples. We conducted two experiments. Initially, we pre-train using the whole generated dataset, then we add a filtering stage using the base classifier. We concluded that filtering removes more noise in the generated samples and achieves better results overall as in Table \ref{table:Filtering}.
\begin{table}[htbp]
\centering
 \resizebox{\linewidth}{!}{
\begin{tabular}{c c c c c}
\toprule
Method   & Sensitivity & Specificity & AUC & harmonic mean\\
\midrule	
All samples  &\textbf{ 93.66} & 89.49 & 97.32  & 91.5  \\
Filtered samples   & 93.09 & \textbf{92.1} & \textbf{98.02} & \textbf{92.59} \\
\bottomrule
\end{tabular}}
\caption{Results on applying DiffuPT with all generated samples and only filtered samples.}
\label{table:Filtering}
\end{table}
\vspace{-5pt}
\subsection{AIROGS Experiments}
Using the AIROGS Dataset, we applied DiffuPT to assess generalization. We deducted that DiffuPT could improve the outcomes of the baseline classifier as seen in Table \ref{table:AIROGs_Test}. Here the baseline classifier is the best performing using a weighted sampler.
We conclude that by our results on both our national dataset and AIROGS,  DiffuPT can generalize on different glaucoma detection datasets.
\begin{table}[htbp]
\centering
 \resizebox{\linewidth}{!}{
\begin{tabular}{c c c c c}
\toprule
Method   & Sensitivity & Specificity & AUC & harmonic mean\\
\midrule	
Weighted Sampler   & 91.66 & 96.97 & 98.84 & 94.24\\
DiffuPT   &\textbf{ 94.44} & \textbf{97.19} & \textbf{98.98} & \textbf{95.8} \\
\bottomrule
Weighted Sampler    & 87.36& \textbf{96.47}& 97.38& 91.69\\
DiffuPT   & \textbf{89.08}& 95.7& \textbf{97.58}& \textbf{92.27}\\
\bottomrule
\end{tabular}}
\caption{AIROGS results - Validation (top) and Test (bottom).}
\label{table:AIROGs_Test}
\end{table}



\vspace{-10pt}
\section{Conclusion}
This paper proposes a new method to use generative models in computer vision tasks such as classification to overcome class imbalance in datasets. Our proposed method surpassed existing methods such as standard augmentation, weighted sampling, weighted loss, and multistage training. Moreover, it proved robustness across two datasets. We also compare several generation models (GANs and Diffusion models). Our final framework requires no \textit{man in the loop} intervention and improves the classifier's overall performance. One drawback might be the overhead of the generation process.

\section{Acknowledgment}
The authors gratefully acknowledge the Applied Innovation Center (AIC) of the Egyptian Ministry of Communication and Information Technology for funding the research presented in this paper. Special thanks are extended to Dr. Ahmed Tantawy, Director of AIC, for his initiative in launching this project.
{\small
\bibliographystyle{ieee_fullname}
\bibliography{Paper}
}

\appendix
\section{Conditional Vs Unconditional generation}
\begin{figure}[htbp]
  \centering
        \centering
        \includegraphics[width= 0.7\linewidth]{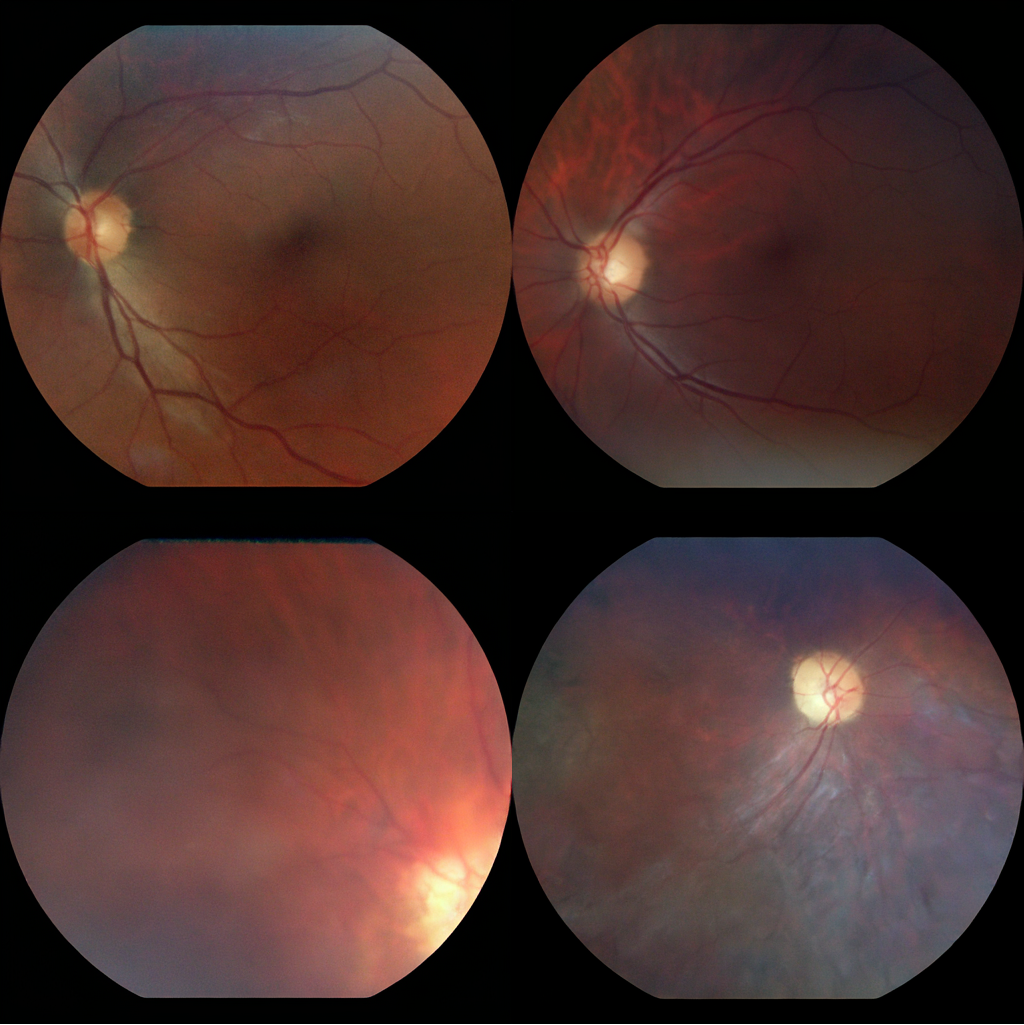} 
        \newline
        \caption{The top images are from conditional LDM and the bottom images show the unconditional generation.}
        \label{fig:Image comparison2}
\end{figure}

Two generative techniques were considered to generate glaucoma samples. First, we tried to generate glaucoma images only using unconditional generation. This can be achieved by training DM on glaucoma images only. However, sample quality was substandard due to the lack of enough glaucoma samples, thus we tried to generate samples conditioned on the class label. This can be achieved by training the DM on both glaucoma and non-glaucoma images. The class conditioning can be achieved during sampling using the equation \ref{eq:no classifier guidance}. During sampling, we specify the target class and unconditional guidance scale $w = 3.0$ to control the quality of generation. The discrepancies in the results can be further noticed in the quality of the image produced. This can be clearly shown in Figure \ref{fig:Image comparison2}.
\begin{equation}
    \tilde{\epsilon}_\theta(z_t, c) = (1 + w)\epsilon_\theta(z_t, c) - w\epsilon_\theta(z_t , \emptyset)
    \label{eq:no classifier guidance}
\end{equation}
\section{Failed Attempts}
\begin{figure}[htbp]
  \centering
        \centering
        \includegraphics[width= 0.9\linewidth]{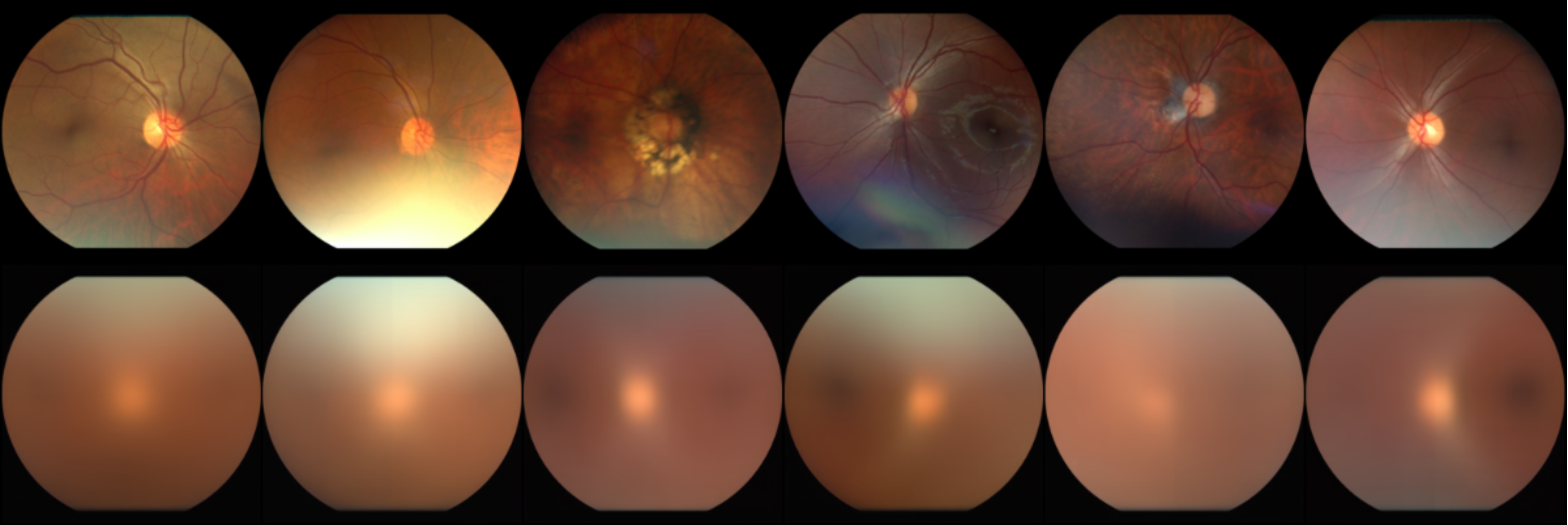} 
        \newline
        \caption{The top images are reconstruction images. The bottom images shows the generated samples.}
        \label{fig:VAE output}
\end{figure}

We show failed attempts for fundus image generation. We attempted to generate images conditionally using Conditional VAE \cite{sohn2015learning}. We tried different encoder-decoder architectures but none was successful. We show images of Conditional VAE in Figure \ref{fig:VAE output}.

\begin{figure*}[th]
  \centering
        \centering
        \includegraphics[width= 1\linewidth]{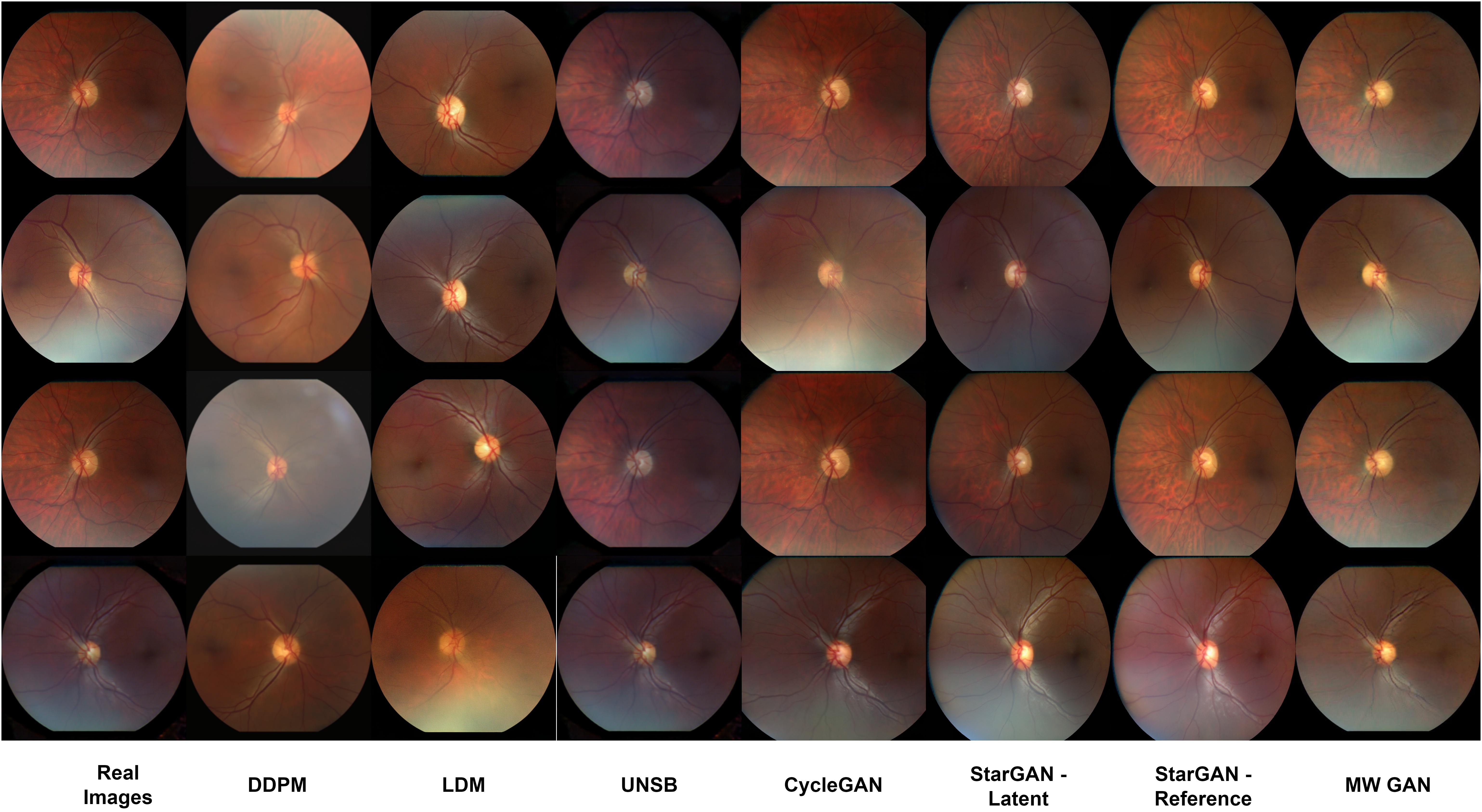} 
        \newline
        \caption{More Generated Samples.}
        \label{fig:Image comparison}
\end{figure*}

\section{Generation}

We show more samples of the generation in Figure \ref{fig:Image comparison}, MWGAN contains more artifacts than other models. LDM samples show the best quality and clearest images.
\section{Results analysis}
\label{sec:confusion}

\begin{figure*}[htbp]
  \centering
        \centering
        \includegraphics[width= 1\linewidth]{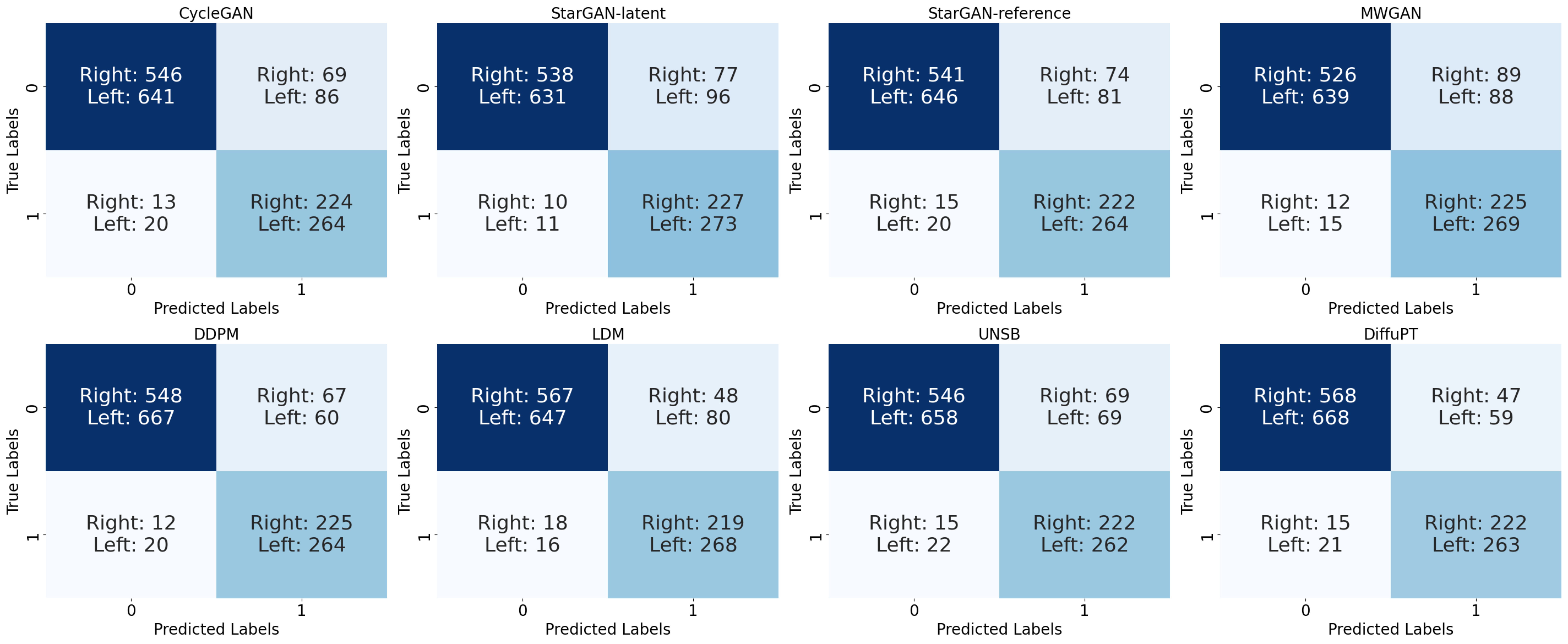} 
        \newline
        \caption{Confusion Matrices showing the right eyes separately from the left eyes.}
        \label{fig:confusion}
\end{figure*}

We show confusion matrices for normal Training, augmentation method, and diffuPT. The confusion matrices in Figure \ref{fig:confusion} show that diffuPT doesn't favor one eye over the other (left and right). One general trend noticed that there wasn't any clear bias of any generative model for the left eye over the right.

\end{document}


\title{DiffuPT: Class Imbalance Mitigation for Glaucoma Detection via Diffusion Based Generation and Model Pretraining \\  Supplementary Materials}

\maketitle


\appendix
\section{Conditional Vs Unconditional generation}
Two generative techniques were considered to generate glaucoma samples. First, we tried to generate glaucoma images only using unconditional generation. This can be achieved by training DM on glaucoma images only. However, sample quality was substandard due to the lack of enough glaucoma samples, thus we tried to generate samples conditioned on the class label. This can be achieved by training the DM on both glaucoma and non-glaucoma images. The class conditioning can be achieved during sampling using the equation \ref{eq:no classifier guidance}. During sampling, we specify the target class and unconditional guidance scale $w = 3.0$ to control the quality of generation. The discrepancies in the results can be further noticed in the quality of the image produced. This can be clearly shown in Figure \ref{fig:Image comparison}.
\begin{equation}
    \tilde{\epsilon}_\theta(z_t, c) = (1 + w)\epsilon_\theta(z_t, c) - w\epsilon_\theta(z_t , \emptyset)
    \label{eq:no classifier guidance}
\end{equation}
\begin{figure}[h]
  \centering
        \centering
        \includegraphics[width= 0.7\linewidth]{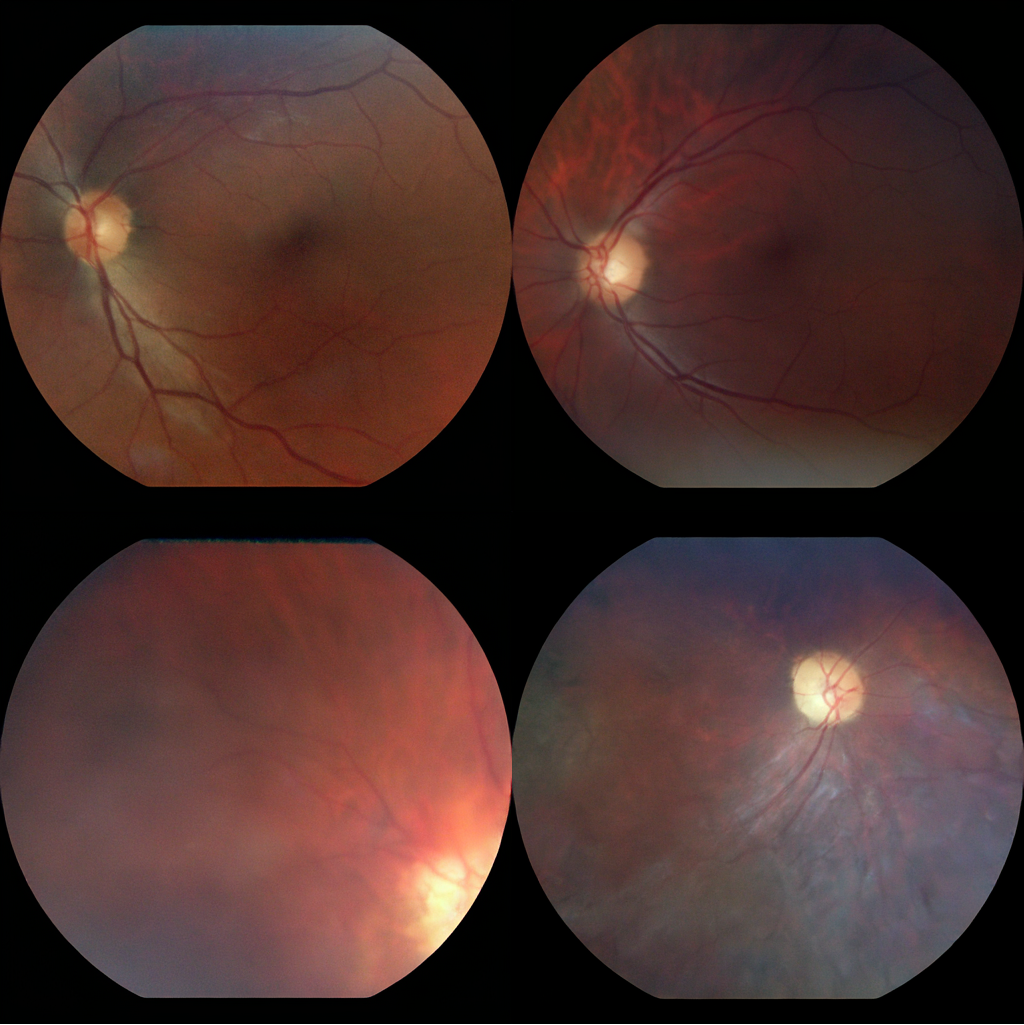} 
        \newline
        \caption{The top images are from conditional LDM and the bottom images show the unconditional generation.}
        \label{fig:Image comparison}
\end{figure}
\section{Failed Attempts}
\begin{figure}[h]
  \centering
        \centering
        \includegraphics[width= 0.9\linewidth]{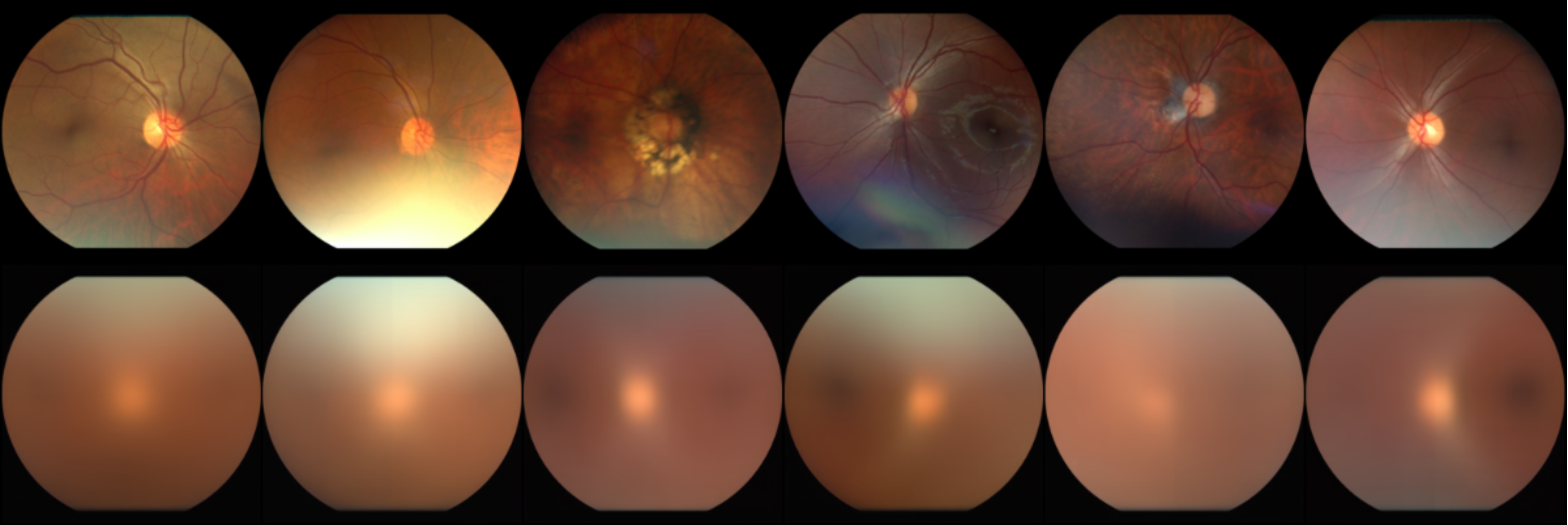} 
        \newline
        \caption{The top images are reconstruction images. The bottom images shows the generated samples.}
        \label{fig:VAE output}
\end{figure}

We show failed attempts for fundus image generation. We attempted to generate images conditionally using Conditional VAE \cite{sohn2015learning}. We tried different encoder-decoder architectures but none was successful. We show images of Conditional VAE in Figure \ref{fig:VAE output}.

\begin{figure*}[th]
  \centering
        \centering
        \includegraphics[width= 1\linewidth]{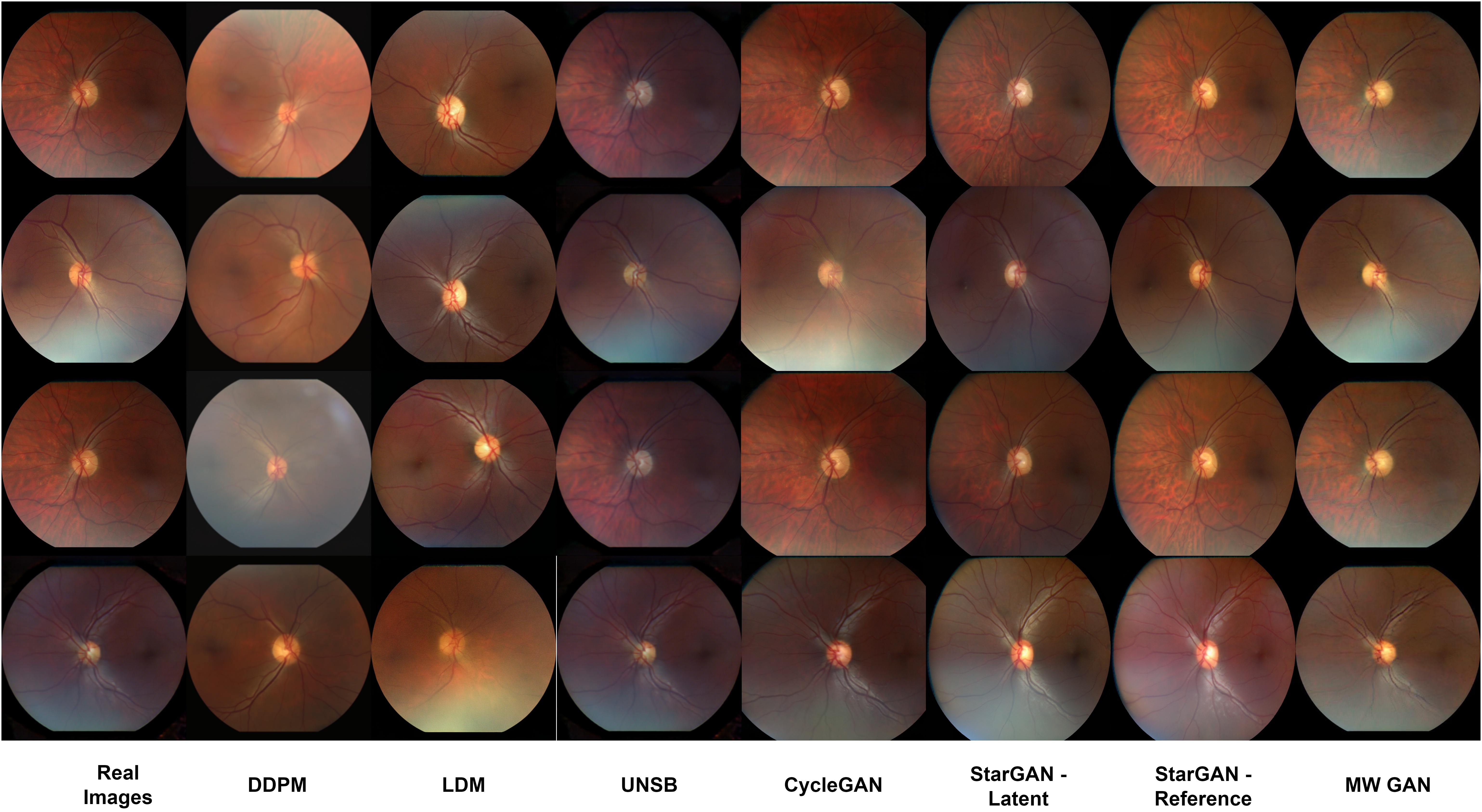} 
        \newline
        \caption{More Generated Samples.}
        \label{fig:Image comparison}
\end{figure*}

\section{Generation}

We show more samples of the generation in Figure \ref{fig:Image comparison}, MWGAN contains more artifacts than other models. LDM samples show the best quality and clearest images.
\section{Results analysis}
\label{sec:confusion}

\begin{figure*}[h]
  \centering
        \centering
        \includegraphics[width= 1\linewidth]{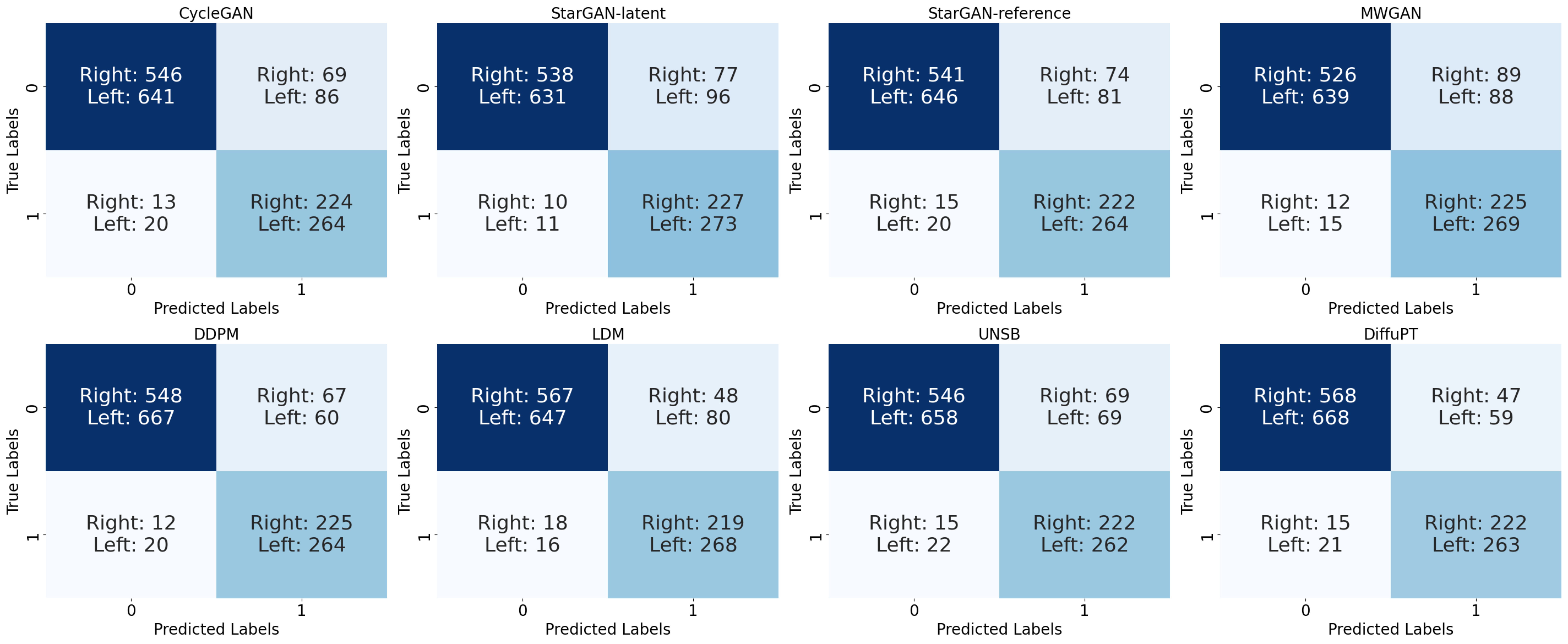} 
        \newline
        \caption{Confusion Matrices showing the right eyes separately from the left eyes.}
        \label{fig:confusion}
\end{figure*}

We show confusion matrices for normal Training, augmentation method, and diffuPT. The confusion matrices in Figure \ref{fig:confusion} \show that diffuPT doesn't favor one eye over the other (left and right). One general trend noticed that there wasn't any clear bias of any generative model for the left eye over the right.

{\small
\bibliographystyle{ieee_fullname}
\bibliography{main}
}